\documentclass[10pt,twocolumn,superscriptaddress,floats,showpacs,nobalancelastpage,longbibliography,prb]{revtex4-2}
\usepackage[latin9,utf8]{inputenc}
\usepackage{color}
\usepackage{babel}
\usepackage{verbatim}
\usepackage{bm}
\usepackage{amsmath}
\usepackage{amssymb, mathtools}
\usepackage{graphicx}
\usepackage{indentfirst}
\usepackage[unicode=true,pdfusetitle,
bookmarks=true,bookmarksnumbered=false,bookmarksopen=false,
breaklinks=false,pdfborder={0 0 0},pdfborderstyle={},backref=false,colorlinks=true]
{hyperref}
\hypersetup{
	citecolor=blue,urlcolor=black}
\usepackage[normalem]{ulem}

\makeatletter

\usepackage{epstopdf}
\usepackage{amsfonts}
\usepackage{mathrsfs}\usepackage{bm}\usepackage{appendix}
\usepackage{txfonts}
\usepackage{float}
\usepackage{epstopdf}
\usepackage{ulem}

\definecolor{OliveGreen}{cmyk}{0.64, 0, 0.95, 0.40}

\begin{document}

\title{Fractional Chern insulator candidate in twisted bilayer checkboard lattice} 

\author{Jia-Zheng Ma}
\affiliation{Guangdong Provincial Key Laboratory of Magnetoelectric Physics and Devices, State Key Laboratory of Optoelectronic Materials and Technologies, Center for Neutron Science and Technology, School of Physics, Sun Yat-sen University, Guangzhou, 510275, China}

\author{Rui-Zhen Huang}
\affiliation{Department of Physics and Astronomy, University of Ghent, 9000 Ghent, Belgium}

\author{Guo-Yi Zhu}
\affiliation{Institute for Theoretical Physics, University of Cologne, Z\"ulpicher Straße 77, 50937 Cologne, Germany}

\author{Ji-Yao Chen}
\email{chenjiy3@mail.sysu.edu.cn}
\affiliation{Guangdong Provincial Key Laboratory of Magnetoelectric Physics and Devices, State Key Laboratory of Optoelectronic Materials and Technologies, Center for Neutron Science and Technology, School of Physics, Sun Yat-sen University, Guangzhou, 510275, China}

\author{Dao-Xin Yao}
\email{yaodaox@mail.sysu.edu.cn}
\affiliation{Guangdong Provincial Key Laboratory of Magnetoelectric Physics and Devices, State Key Laboratory of Optoelectronic Materials and Technologies, Center for Neutron Science and Technology, School of Physics, Sun Yat-sen University, Guangzhou, 510275, China}

\begin{abstract}

We investigate a fractional Chern insulator (FCI) candidate arising from Moiré bands with higher Chern number $C=2$ on a magic angle twisted bilayer checkerboard lattice (MATBCB). There are two nearly flat low lying bands in the single particle energy spectrum under the first magic angle $\phi\approx 1.608^{\circ}$ and chiral limit. We find MATBCB hosts a nearly uniform Berry curvature distribution and exhibits tiny violation of quantum geometric trace condition in the first moir\'e Brillourin Zone (mBZ), indicating that there is a nearly ideal quantum geometry in MATBCB in single particle level. Turning on projected Coulomb interactions, we perform exact diagonalization and find a ten-fold ground state quasi-degeneracy in many body energy spectrum with filling fraction $\nu=1/5$. The ten-fold quasi-degenrate ground states further show spectra flow under flux pumping. By diagnosing the particle entanglement spectrum (PES) of the ground states, we obtain a clear PES gap and quasi-hole state counting consistent with Halperin spin singlet generalized Pauli principle, suggesting that a $C=2$ abelian fractional Chern insulator is realized in this system.

\end{abstract}

\maketitle

\date{\today}

\setcounter{equation}{0}
\setcounter{figure}{0}
\setcounter{table}{0}
\makeatletter

\section{Introduction}
\label{sec:Intro}

Fractional Chern insulator (FCI) is a lattice enriched version of fractional quantum hall effect (FQHE). It is believed to be realizable under weak magnetic field or even with zero external magnetic field, and has attracted many interest in both theoretical and experimental communities. An important potential application of FCI is topological quantum computation~\cite{Nayak2005prl,DasSarma2008rmp}. The existence of FCI phase was predicted by theory decades ago~\cite{XiaoGangWen2011prl,Neupert2011prl,KaiSun2011prl,Bernevig2011prx,DNSheng2011natcomm,XiaoLiangQi2011prl}. Recently, several experiments have realized the fractional quantum anomalous hall effect (FQAHE) in material system including twisted $\rm{MoTe_{2}}$ system~\cite{JiaQiCai2023Nature,Eric2023Nature,TingXinLi2023prx} and rhombohedral pentalayer graphene (R5G)~\cite{LongJu2024nature}. FQAHE can be considered as a phenomenon in zero magnetic field FCI phase with fractional anomalous hall conductivity. Twisted $\rm{MoTe_{2}}$ and R5G become better system than twisted bilayer graphene (TBG) in realizing FCI phase, since the FCI phase in TBG needs weak external magnetic field to make it more stable~\cite{YongLongXie2021Nature}.

These experiments have raised a great interest among theorists. Several theories have been proposed to interpret the FQAHE (zero magnetic field FCI) in $\rm{MoTe_{2}}$~\cite{DiXiao2024prl,LiangFu2023arxiv,Bernevig2023arxiv,Regnault2023arxiv,WangYao2024arxiv,Regnault2024Commphys,WeiZhu2024arxiv,sharma2024arxiv,FengChengWu2024arxiv,Santos2024arxiv} and R5G~\cite{Senthil2024arxiv,JianPengLiu2023arxiv,Senthil2023arxiv,TiXuanTan2024arxiv,TaigeWang2023arxiv,YaHuiZhang2023arxiv,JiaBinYu2023arxiv}. Besides gapped phases, gapless phases beyond FQAHE, e.g., the composite fermi liquid phase, have also been found~\cite{JunKaiDong2023prl}. However, the experiments mentioned above all seem to focus on fractional filling $C=1$ Chern band which has well known lowest Landau level (LLL) correspondence. When the topological flat band host a higher Chern number, its potential FCI phase will not have a naive LLL correspondence~\cite{Regnault2013prb,Regnault2014prb,ZhaoLiu2012prl,ZhaoLiu2022prl,YiFeiWang2012prb}. What is more attracting, it is possible to realize non-abelian FCI phase~\cite{XiaoBoLu2024arxiv,Regnault2013prb,DiXiao2024arxiv,XueYangSong2024arxiv,YangZhang2024arxiv,LiangFu2024arxiv,Emil2024arxiv,WeiZhu2024arxiv,JieWang2024arxiv,AiLeiHe2020prb,Barkeshli2024arxiv,huiliu2024arxiv,JiaQiCai2024arxiv,TingXinLi2024arxiv} in fractionally filled flat band with higher Chern number, which could be a building block for fault-tolerant quantum computation.

Motivated by above progress, it is valuable to search for higher Chern number FCI phase in real material or artificial system. In this manuscript, we theoretically study a $C=2$ abelian FCI candidate in magic angle twisted bilayer checkerboard (MATBCB) lattice model~\cite{HongYao2022prr} with fractional filling $\nu=1/5$ of the lowest single particle conductive band. By using exact diagonalization (ED) in momentum space, we find there is a ten-fold ground state quasi-degeneracy in the many-body energy spectrum and the spectrum flow under adiabatic flux inserting. These ten-fold low energy states are the topological ground state candidates for putative $C=2$ FCI phase.  From the perspective of many-body entanglement, we study the particle entanglement spectrum (PES) of these ten-fold ground states density matrix.  We find that the quasi-hole counting under PES gap obeys the Halperin spin singlet generalized Pauli principle (GPP), which is a smoking gun evidence for the existence of $C=2$ FCI phase. This higher Chern number FCI phase may be realized in artificial quantum simulator such as Rydberg atom array~\cite{YangZhao2022arxiv}, twisted optical lattice~\cite{Cirac2019pra,ZheYuShi2024arxiv,YoujinDeng2024arxiv,LiangHe2024arxiv}, circuit QED system~\cite{JianWeiPan2024Science}. Up to now, we have noticed that several works have studied TBCB flat band~\cite{HongYao2022prr,XiaoHanWan2023arxiv}, or similar system like twisted square lattice~\cite{bernevigChew2022prb,Fukui2022prb,Paredes2024arxiv,ZhuXiLuo2024arxiv}, twisted FeSe~\cite{Vafek2023SciPostPhys}, strained moir\'e lattice with quadratic dispersion~\cite{XiaoHanWan2023prl}, and square lattice moir\'e heterostructure~\cite{LeDeXian2024arxiv}.

This manuscript is organized as follows: In section ~\ref{sec:single_body_TBCB}, we briefly review the single particle Bistrizer-MacDonald (BM) like model of TBCB and derive the topologically non-trivial two lowest flat bands with higher Chern number $C=\pm 2$. In section~\ref{sec:QGT}, we examine the single particle quantum geometry indicator for MATBCB flat band and check if it is suitable for MATBCB to host FCI phase. Then in section~\ref{sec:ED_and_spec_flow}, we turn to the flat band projected interacting many-body model of MATBCB with filling fraction $\nu=1/5$ to reveal the potentially topological many-body ground states and their spectrum flow under adiabatic flux inserting. In section~\ref{sec:PES}, we further present the PES of putative topological ground state density matrix and show the quasi-hole statistics. The quasi-hole counting obeys the Halperin spin singlet GPP, indicating the phase in TBCB under magic angle and chiral limit is a FCI instead of a CDW or other competing phases. In section~\ref{sec:disc}, we conclude and discuss further directions with large-scale numerical approaches for hunting FCI in moir\'e system with higher Chern number.

\begin{figure*}[htb]
\centering
\includegraphics[width=1.85\columnwidth]{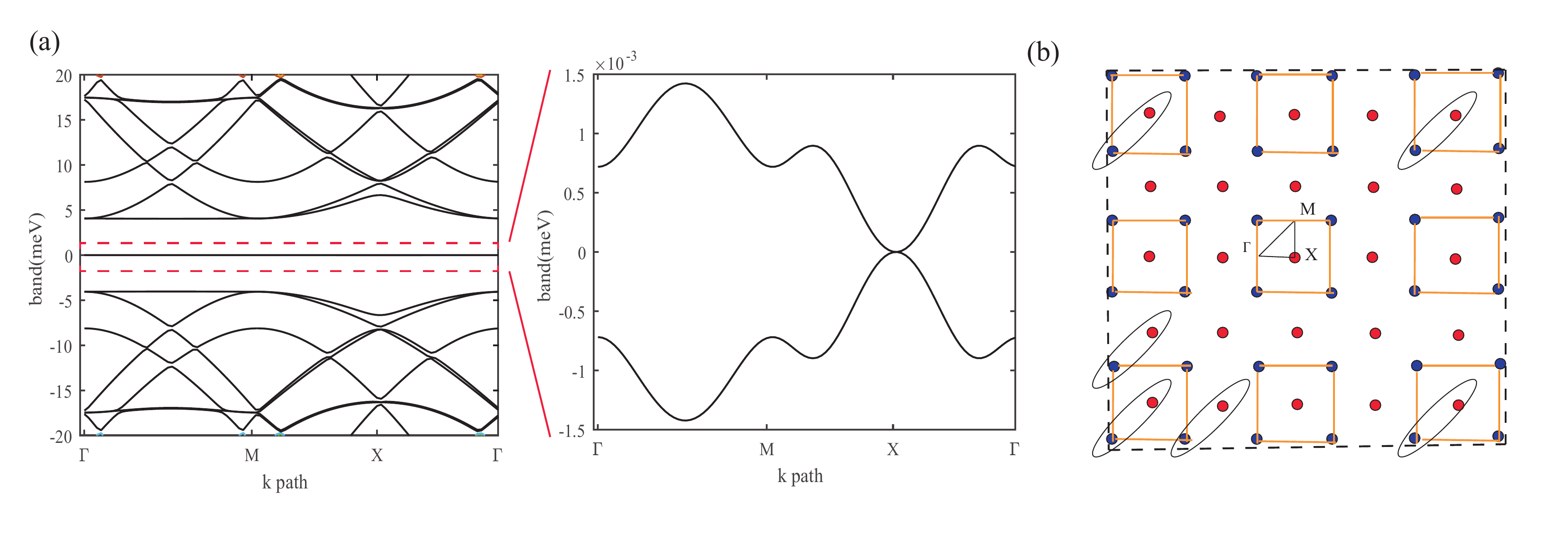} 
\caption{(a) The single particle Moir\'e band structure along high symmetry path in the first mBZ and the zoomed in figure of the lowest two flat bands. One can see that the bandwidth at the first magic angle $\phi\approx 1.608^{o}$ and chiral limit $\kappa=w_{AA}/w_{AB}=0$ is about $1.5\times 10^{-3} {\rm meV}$. There is a quadratic band touching at X point which reflects the nature of tight binding checkboard lattice model dispersion when $\Delta=0$. The number of k points in the high symmetry path is 100 and the size of quadratic node array is $25\times 25$ in single particle calculation. (b) The scheme of quadratic gapless node array in momentum space for TBCB, which is similar to Dirac cone in TBG. Blue ones represent the top layer nodes and red ones are bottom layer nodes. The yellow square is one of the choice of the first mBZ. Moir\'e tunneling will occur between each pair of the nearest nodes. The elliptic can be considered as the minimal unit in this array.}
\label{fig:singlebody_band}
\end{figure*}

\begin{figure*}[htb]
\includegraphics[width=1.8\columnwidth]{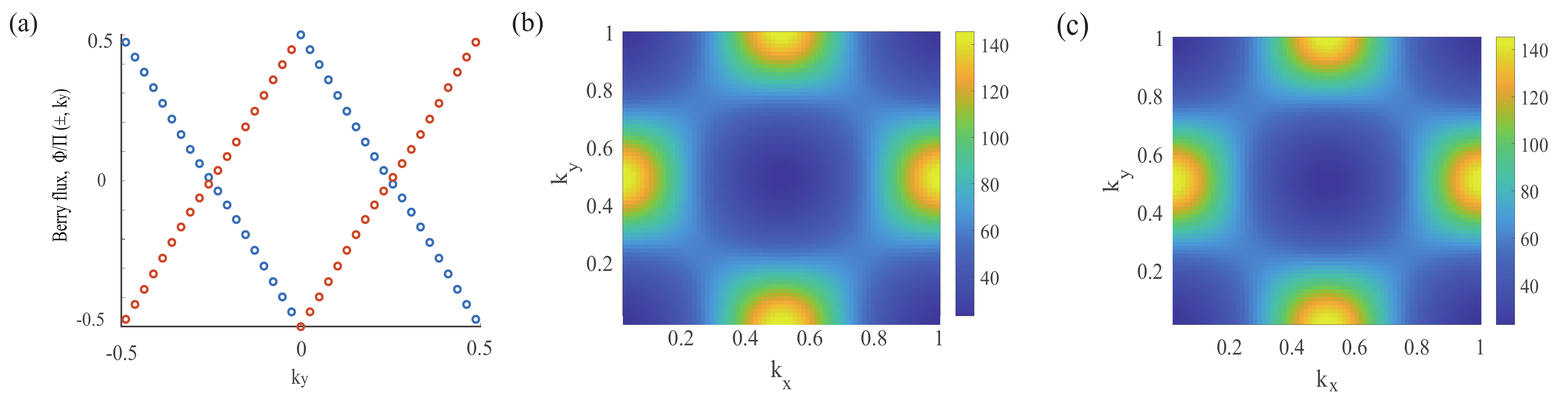}
\caption{(a) The Wilson loop winding of the lowest two flat bands near half filling. Since the single particle band is gapless at X, the Wilson loop is a $2\times 2$ matrix. The windings are corresponding to the two eigenvalues of the Wilson loop. One can read out the Chern number $C=\pm 2$.  From the variation speed of the winding, one can believe that the Berry curvature distribution is uniform enough to host a FCI phase. (b) The Berry curvature given by Eq.~\eqref{eq:Wilson_LGT}. The k point density in the first mBZ is $74\times 74$. (c) The trace of quantum metric in the first mBZ using the same k point density. It almost has the same distribution with Berry curvature, which implies the single particle flat band Bloch function is close to a holomorphic function about $k=k_{x}+ik_{y}$ and ideal trace condition violation. The range of momentum $k_{x},k_{y}$ are the same $[0,2\pi]$ for (b) and (c).
}
\label{fig:Berry_curvature}
\end{figure*}

\section{Single particle moir\'e band structure of MATBCB}
\label{sec:single_body_TBCB}


In this section we solve the Moir\'e Hamiltonian (free fermion hopping in momentum space) of MATBCB and show the flat band structure, as well as the non-trivial Chern number. 
To clarify, in the following we will use lower case $k$ to label the single particle momentum while the upper case $K$ denotes the many body momentum or total momentum. The two dimensional single particle wave vector is defined as $\textbf{k}=(k_{x}, k_{y})$. The Greek symbol $\kappa=w_{AA}/w_{AB}$ represents chiral ratio to be introduced later.
The Moir\'e Hamiltonian for a given momentum $H(k_x, k_y)$ expressed in reciprocal lattice basis is shown in Eq.~\eqref{eq:BM}:

    \begin{equation}
	\begin{split}	
& H=\sum_{k,q_{i,b},q_{i,t}}h(k,q_{i,b},q_{i,t})
 +\\ &\sum_{k,\langle q_{j,t},q_{i,b}\rangle}
	\left(\begin{array}{cc} c^{\dagger}_{k,q_{j,t}} & c^{\dagger}_{k,q_{i,b}}
	\end{array}\right)
	\left(\begin{array}{cc} 0 & T_{q_{j,t},q_{i,b}}\\
	T_{q_{j,t},q_{i,b}}^{\dagger} & 0 \end{array}\right)\left(\begin{array}{cc} c_{k,q_{j,t}} \\ c_{k,q_{i,b}}
	\end{array}\right) ,~\label{eq:BM}
	\end{split}
 \end{equation}
where we have denoted  
 \begin{equation}
	\begin{aligned}
	&h(k,q_{b},q_{t})=\\ 
 &\left(\begin{array}{cc} c^{\dagger}_{k,q_{t}} & c^{\dagger}_{k,q_{b}}
	\end{array}\right)
	\left(\begin{array}{cc} H_{0}^{\phi/2}(k-q_{t}) & T_{q_{t},q_{b}}\\
	T_{q_{t},q_{b}}^{\dagger} & H_{0}^{-\phi/2}(k-q_{b})\end{array}\right)\left(\begin{array}{cc} c_{k,q_{t}} \\ c_{k,q_{b}}
	\end{array}\right) ,~\label{eq:moire}\\ 
    \end{aligned}
    \end{equation}
with    
    \begin{equation}
	\begin{split}
	&H_{0}^{\phi}(k)=2t'\cos(k_{x}-k_{y})(\cos\phi\sigma_{y}+\sin\phi\sigma_{x})+ \\
 & 4t\cos(\frac{k_{x}+\pi}{2})\cos(\frac{k_{y}+\pi}{2})(\cos\phi\sigma_{x}-\sin\phi\sigma_{y})+\Delta \sigma_{z} ,~\label{eq:monolayer}\\
    \end{split}
    \end{equation}
and
 \begin{equation}
	\begin{split}
	&T_{q_{t},q_{b}}=(w_{AA}I+w_{AB}\sigma_{x})[\delta(q^{x}_{t}=q^{x}_{b}+\frac{k_{\phi}}{\sqrt{2}})\delta(q^{y}_{t}=q^{y}_{b}+\frac{k_{\phi}}{\sqrt{2}})\\
 &+\delta(q^{x}_{t}=q^{x}_{b}-\frac{k_{\phi}}{\sqrt{2}})\delta(q^{y}_{t}=q^{y}_{b}-\frac{k_{\phi}}{\sqrt{2}})]+\\
	&(w_{AA}I-w_{AB}\sigma_{x})[\delta(q^{x}_{t}=q^{x}_{b}+\frac{k_{\phi}}{\sqrt{2}})\delta(q^{y}_{t}=q^{y}_{b}-\frac{k_{\phi}}{\sqrt{2}})\\
 &+ \delta(q^{x}_{t}=q^{x}_{b}-\frac{k_{\phi}}{\sqrt{2}})\delta(q^{y}_{t}=q^{y}_{b}+\frac{k_{\phi}}{\sqrt{2}})]~\label{eq:tunnel} ,\\ 
 \end{split}
    \end{equation}

Exact diagonalizing this Hamiltonian gives rise to the Moire band structure $\epsilon(k_x, k_y)$ and the Bloch orbitals $u(k_x, k_y)$. The band structure is shown in Fig.~\ref{fig:singlebody_band}.

Here $H_{0}$ is the Hamiltonian for monolayer checkboard lattice with quadratic dispersion. $T_{q_{t}, q_{b}}$ is the interlayer coupling between quadratic nodes $q_{t},q_{b}$. $T_{q_{t},q_{b}}$ includes intra-mBZ term and inter-mBZ term. $h(k,q_{b},q_{t})$ is the Hamiltonian within mBZ. The TBCB parameters include intra-sublattice hopping $t'$ and inter-sublattice hopping $t$, interlayer coupling in AA region $w_{AA}$ and the AB region counterpart $w_{AB}$. In Eq.~\eqref{eq:tunnel}, $k_{\phi}=\frac{2\sqrt{2}\pi}{a}\sin(\phi/2)$ is the distance between adjacent quadratic points from each layer in momentum space and $a$ is the lattice constant for monolayer. Here we restrict the parameters at the first magic angle $\phi=1.608^{\circ}$, chiral limit $t'=t/2=500\ {\rm meV}, w_{AB}=2.05\ {\rm meV}, w_{AA}=0$ as suggested in Ref.~\cite{HongYao2022prr}. In addition, we have added a sublattice mass term $\Delta \sigma_{z}$ compared with~\cite{HongYao2022prr} to make many body computation simpler in the next section. The representation space of flat band Bloch orbital is the direct product of reciprocal point,  sublattice and layer space. As shown in Fig.~\ref{fig:singlebody_band}, red points represent the reciprocal lattice of bottom layer and blue points are the counterpart for top layer.

To elucidate the topological nature of the flat band, we evaluate the non-contractible Wilson loop circumventing the mBZ in $x$ direction, defining $|u_{\alpha}(\bf{k})\rangle$ as the periodic  part of bloch function at band $\alpha$. One has the following overlap matrix and Wilson loop: 
	\begin{equation}
	\begin{split} \label{eq:winding}
	&[O(k_{x},k_{y})]_{\alpha\beta}=\langle u_{\alpha}(k_{x},k_{y})|u_{\beta}(k_{x}+dk_{x},k_{y})\rangle,\quad \\
 &dk_{x}=\frac{2\pi }{N_{x}},\quad \alpha,\beta=1,2. \\
	&W(k_{y})=\left[\prod_{n_{x}=1}^{N_{x}-1}O\left(k_{x}=\frac{2\pi}{N_{x}}(n_{x}-\frac{N_{x}+1}{2}),k_{y}\right)\right]O'(k_{y}),\\
	&[O'(k_{y})]_{\alpha\beta}=\sum_{G_{x},G_{y},\tau,\mu}\langle u_{\alpha}(k_{x}=\frac{\pi(N_{x}-1)}{N_{x}},k_{y})|G_x,G_y,\tau,\mu\rangle \\ 
 &\langle G_x-2\pi,G_y,\tau,\mu|u_{\beta}(k_{x}=-\frac{\pi(N_{x}-1)}{N_{x}},k_{y})\rangle.
	\end{split}
	\end{equation}

From which we extract the global Berry flux threaded into the compactified mBZ towards the $k_{y}$ direction. The Berry flux for the quasi-degenerate two flat bands are shown in Fig.~\ref{fig:Berry_curvature}(a), which means the flat bands carry Chern numbers $C=\pm 2$ respectively.

The eigenvalues for matrix $W(k_{y})$ in Eq.~\eqref{eq:winding} are just the winding phases of the polarization. The eigenfunction of Hamiltonian Eq.~\eqref{eq:BM} under reciprocal point, sublattice, layer direct product basis is labeled by $\langle G_x,G_y,\tau,\mu|u_{\alpha}(k_{x},k_{y})\rangle$, and $G=(G_x,G_y)$ is the reciprocal vector. $\alpha,\tau,\mu$ represent band, sublattice and layer respectively. The complete relation is  $\sum_{G_x,G_y,\tau,\mu}|G_x,G_y,\tau,\mu\rangle \langle G_x,G_y,\tau,\mu|=I$. When the winding link goes across the boundary of the first mBZ, one should shift the component of the Bloch wavefunction by a primitive reciprocal vector in x direction to ensure that the overlap $O'$ is still a first-order infinitesimal matrix. It corresponds to inserting the embedding matrix~\cite{HongYao2022prr,ZhiDaSong2021prbII,ZhiDaSong2021prbIV} like Eq.~\eqref{eq:winding}. The k space embedding matrix has its real space counterpart in Ref.~\cite{JianKang2020prb}. For a given $k_y$, in principle one can start from arbitrary $k_x$ to get the winding $W'(k_{x},k_{y})$ as long as inserting embedding matrix when crossing the mBZ boundary. One can easily prove that $W(k_y)$ and $W'(k_x,k_y)$ share the trace and determinant. So the eigenvalues for $W(k_y)$ and $W'(k_x,k_y)$ are the same for arbitrary $k_x$. However, the eigenvectors for  $W(k_y)$ and $W'(k_x,k_y)$ are generally different. These eigenvectors can be related by wilson line like Ref.~\cite{JianKang2020prb}.

Importantly, when $N_{x}\to\infty, dk_x\to 0$, $O(k_x,k_y)$ and $W(k_y)$ are unitary matrices. Nevertheless in realistic numerical implementation, $N_x$ is finite so $O(k_x,k_y)$ will be a non-unitary matrix.  
In order to get unitary $W(k_y)$ and its pure phase eigenvalues, one has to extract the unitary part of $O(k_x,k_y)$. Available strategy can be singular value decomposition (SVD)~\cite{JianKang2020prb}
	\begin{equation}
	\begin{split}
	&O(\textbf{k})=U(\textbf{k})S(\textbf{k})V^{\dagger}(\textbf{k}),\quad \tilde{O}(\textbf{k})=U(\textbf{k})V^{\dagger}(\textbf{k}),\\
 \end{split}
 \end{equation}
 or polar decomposition~\cite{Zaletel2020prb}:
 \begin{equation}
 \begin{split}
	&O(\textbf{k})=P(\textbf{k})R(\textbf{k}),\quad R(\textbf{k})=\sqrt{O^{\dagger}(\textbf{k})O(\textbf{k})},\quad 
\\
&P(\textbf{k})=\tilde{O}
    (\textbf{k})=O(\textbf{k})R^{-1}(\textbf{k})
    =O^{1/2}(\textbf{k})[O^{\dagger}(\textbf{k})]^{-1/2}.
\end{split}
\end{equation}

Here we denote the unitary part of $O(k_x,k_y)$ as $\tilde{O}(k_x,k_y)$. Accordingly, $\tilde{W}(k_y)$ will also be a unitary matrix. With these tricks, one can go back and inspect the single particle energy spectrum and Wilson loop. In Fig. ~\ref{fig:singlebody_band}, we set the sublattice staggered potential $\Delta=0$. And we find that the lowest quadratic flat bands in MATBCB are about $1.5\times 10^{-3}\quad meV$ wide. From the Wilson loop winding of flat bands shown in \ref{fig:Berry_curvature}(a), one can see that there is almost uniform winding over the mBZ. It is reminiscent that MATBCB may satisfy the ideal quantum geometry condition well, which is suitable for hosting FCI phase.

Up to now we have a preliminary picture of MATBCB in single particle level including single particle band and Wilson loop winding. In the next section, we will use quantum geometry indicators to inspect if MATBCB is favor to host FCI phase more carefully.


\section{Quantum geometry indicators for fractional Chern insulator}
\label{sec:QGT}

In this section we compute the distribution of Berry curvature and the quantum geometry tensor in mBZ, to inspect the quantitative difference of our flat band Bloch orbitals from the lowest Landau levels (LLL).  The quantum geometry indicators for FCI phase in single particle level include~\cite{parker2021arxiv,JieWang2024arxiv}:
\begin{enumerate}
\item The bandwidth need to be narrow enough to suppress the kinetic term. It should be much more narrow than single particle excitation gap and interaction energy scale.

\item The standard deviation of Berry curvature over the first mBZ $\sigma[\eta]=[\frac{1}{2\pi}\int d^{2}k(F(\textbf{k})-\bar{F})^{2}]^{1/2}$ should be as small as possible. Here $F(\textbf{k})$ is Berry curvature while $\bar{F}=2\pi C/A$ represents averaged Berry curvature.  C is  Chern number and A is the area of mBZ. The uniform Berry curvature is good to host FCI which corresponds to LLL.

\item The trace condition violation $T[\eta]=\frac{1}{2\pi}\int d^{2}k (Tr(g)-|F|)$. g is quantum metric (Fubini-Study metric) which will be introduced in details later. $T[\eta]$ measures the deviation between flat band bloch wave function and holomorphic function about $k=k_{x}+ik_{y}$. To host the FCI phase, this violation is also required to be as small as possible. 
\end{enumerate}

In fact, there is still some subtlety when applying quantum geometry indicator to higher Chern number band,  for which a recent generalization can be found in Ref.~\cite{JieWang2023prr,JieWang2024arxiv}. Here for simplicity, out strategy is to transform the higher Chern number problem to familiar $C=1$ case. Fortunately, several recent works have shown that ideal higher Chern band can be decomposed into a set of $C=1$ ideal Chern bands~\cite{junkaidong2023prr,Patrick2022prl} under certain condition. Specially, when the Berry curvature of $C>1$ Chern band satisfies $F(\textbf{k})=F(\textbf{k}+\textbf{Q})$ with $\textbf{Q}$ being some translation symmetry broken vector, the set of decomposed $C=1$ ideal Chern bands can form orthogonal basis. That means these $C=1$ bands can be independently filled. So the $C=1$ quantum geometry indicators can be used as usual. In our MATBCB case, $\textbf{Q}=(G_{1}/2,G_{2}/2)$. One can see this point later in detail. A more general FCI indicator may be the so- called vortexability~\cite{Patrick2022arxiv,JunKaiDong2024arxiv}, which we leave for future work.

In our two flat bands subspace, we define the multiband version infinitesimal distance between adjacent Bloch states and the quantum geometry tensor as follows:
\begin{equation}
\begin{split}
&ds^{2}=|\sum_{n}\langle u^{a}_{\textbf{k}}|u^{n}_{\textbf{k}}\rangle\langle u^{n}_{\textbf{k}}|u^{b}_{\textbf{k}}\rangle-\sum_{n}\langle u^{a}_{\textbf{k}+d\textbf{k}}|u^{n}_{\textbf{k}}\rangle\langle u^{n}_{\textbf{k}}|u^{b}_{\textbf{k}+d\textbf{k}}\rangle|\approx
\eta^{ab}_{\mu\nu}dk^{\mu}dk^{\nu},\\
&\eta^{ab}_{\mu\nu}=\langle \partial_{k^{\mu}}u^{a}_{\textbf{k}}|(I-\sum_{n}|u^{n}_{\textbf{k}}\rangle \langle u^{n}_{\textbf{k}}|)|\partial_{k^{\nu}}u^{b}_{\textbf{k}}\rangle=\langle \partial_{k^{\mu}}u^{a}_{\textbf{k}}|(I-P(\textbf{k}))|\partial_{k^{\nu}}u^{b}_{\textbf{k}}\rangle
  \label{eq:QGT1}
	\end{split}
\end{equation}
Here the repeated index $\mu,\nu$ means summing over. $P(\bf{k})$ is the projection operator of two band subspace. The summation of n run over the 2 band subspace. $\mu,\nu$ label the momentum component $\{k_{x},k_{y}\}$. $a,b$ label the band index. $\eta$ is the quantum geometry tensor (QGT)~\cite{parker2021arxiv,JieWang2024arxiv}. Its real part and imaginary part define the quantum metric and Berry curvature respectively, see Eq.~\eqref{eq:QGT2}. We can see that QGT contains more information than Wilson loop or Berry curvature.  
\begin{equation}
	\begin{split}
g^{ab}_{\mu\nu}=\frac{1}{2}(\eta^{ab}_{\mu\nu}+\eta^{ab}_{\nu\mu}), \quad F^{ab}=i\epsilon^{\mu\nu}\eta^{ab}_{\mu\nu}.
  \label{eq:QGT2}
	\end{split}
\end{equation}



For FCI phase, we mainly concern the Berry curvature $F$ and trace of quantum metric $Tr(g)$. In practice, one can use the following Wilson loop formalism Eq.~\eqref{eq:Wilson_LGT} to get $F$, $Tr(g)$ simultaneously and accurately~\cite{JieWang2024arxiv}. For the mBZ (torus $T^{2}$), there is no need to fix the gauge~\cite{JPSJ2005Suzuki}. Caompared with simply using Eq.~\eqref{eq:QGT1}, Eq.~\eqref{eq:Wilson_LGT} can better capture the non-linear character in $F$ and $Tr(g)$ especially when the k point is near the branch cut of Berry phase or the boundary of mBZ. Here we assume $dk_{x}=dk_{y}$.
\begin{equation}
	\begin{split}
&[U_{r}(\textbf{k})]_{ab}=\langle u_{a}(\textbf{k})|u_{b}(\textbf{k}+d\textbf{k}_{r})\rangle, r=x,y.\\
&W(\textbf{k})=[U_{y}(\textbf{k})]^{-1}[U_{x}
(\textbf{k}+dk_{y})]^{-1}U_{y}(\textbf{k}+dk_{x})U_{x}(\textbf{k})\\
&=\exp(-g_{k_{x}k_{x}}dk_{x}^{2}-g_{k_{y}k_{y}}dk_{y}^{2})\exp(iFdk_{x}dk_{y}),\\
&W=USV^{\dagger},\quad \exp(iFdk_{x}dk_{y})= UV^{\dagger},\quad 
S=\exp(-Tr(g)dk_{x}^{2}),\\
&C=\frac{1}{2\pi}\int_{mBZ}d^{2}k Tr(F)=\frac{1}{2\pi i}\int_{mBZ} Tr(\ln(W(\textbf{k})))\\
&=\frac{1}{2\pi i}\int_{mBZ} \ln(\det(W(\textbf{k}))), \\
&F=\frac{\textbf{Im}(\ln(W))}{dk_{x}dk_{y}}, \quad Tr(g)=-\frac{\textbf{Re}(\ln(W))}{dk_{x}^{2}}.
   \label{eq:Wilson_LGT}
	\end{split}
\end{equation}

From Fig.~\ref{fig:Berry_curvature}, one can see that Berry curvature of $C=+2$ band (Wilson eigenvalue) and the trace of quantum metric are consistent, which means the ideal trace condition violation condition is well satisfied. The profile of Berry curvature is similar to the result given by Ref.~\cite{Coleman2024arxiv}, which indicates topological heavy fermion is also a potential competing phase. More importantly, one can see that the Berry curvature host the symmetry  $F(\textbf{k})=F(\textbf{k}+\textbf{Q})$, $\textbf{Q}=(G_{1}/2,G_{2}/2)$. 
That means the $C=2$ band can be orthogonally decomposed into two $C=1$ bands. Therefore the original quantum geometry indicator for $C=1$ case is still valid in our $C=2$ MATBCB system. For MATBCB, we summarize the quantum geometry indicators as follow:
\begin{enumerate}
\item The single body bandwidth is $W=1.5\times 10^{-3}\quad {\rm meV}$. 
\item The standard deviation of Berry curvature is $\sigma[\eta]=6.0579$.
\item The trace condition violation is $T[\eta]=1.6357\times 10^{-7}$. 
\end{enumerate}

Comapared with MATBG~\cite{YongLongXie2021Nature,parker2021arxiv}, MATBCB has more ideal bandwidth and trace condition violation but slightly worse Berry curvature deviation. We note that some recent works have proposed the concept of higher Berry curvature~\cite{XueDaWen2024arxivI,XueDaWen2024arxivII,Ryu2024arxiv}, which may develop to many-body version quantum geometry. Also the trace condition deviation can be generalized to local momentum k version~\cite{BoYang2021prl}. How to apply them to moir\'e system will be an interesting problem.

\begin{figure}[htb]
\includegraphics[width=0.95\columnwidth]{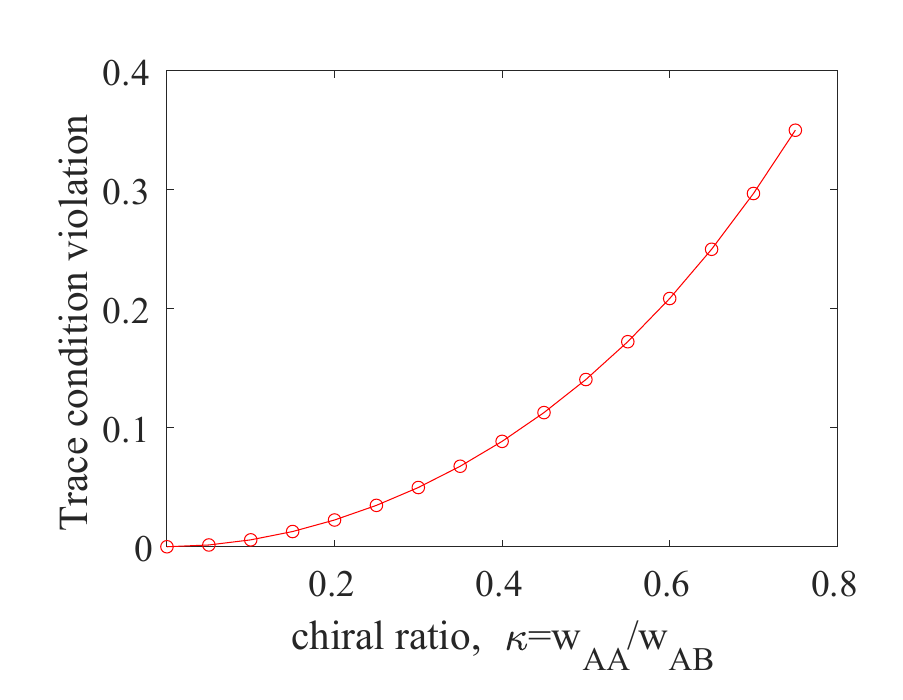}
\caption{The variation of trace condition violation $T[\eta]$ over chiral ratio $\kappa$ under $\Delta=1meV$ case. One can read out that $T[\eta]$ is monotonically increasing with $\kappa$, which indicates that non-zero chiral ratio can induce the instability of putative FCI phase.}
\label{fig:Teta}
\end{figure}

For the non-ideal case, one can consider the non-chiral limit with non-zero $w_{AA}$. The result is shown in Fig.~\ref{fig:Teta}. We inspect the key indicator trace condition violation $T[\eta]$ variation which is directly related to vortexability~\cite{Patrick2022arxiv} about chiral ratio $\kappa$ in single particle level. The $T[\eta]$ is monotonically increasing over $\kappa$. Apparently, chiral limit corresponds to the most ideal quantum geometry for TBCB. 

To sum up for this section, in single particle level, MATBCB is tend to host $C=2$ FCI phase according to quantum geometry indicators and higher Chern number ideal band decomposition formalism.  However, we still need a many-body interpretation of the FCI candidate in MATBCB in the context of many body ground state degeneracy and entanglement. In the next section we will present a many body picture of putative FCI phase in MATBCB.

\section{Many-body energy spectrum and spectrum flow for FCI candidate}
\label{sec:ED_and_spec_flow}

In this section we analyse the corresponding many-body model for MATBCB and solve the low energy spectrum and spectrum flow to see if there is FCI like topological degenerate ground states candidate. For simplicity, we will consider the following projected Coulomb interaction model for topological flat band~\cite{HongYao2022prr}, which is similar to TBG case~\cite{ZhiDaSong2021prbI,ZhiDaSong2021prbII,ZhiDaSong2021prbIII,ZhiDaSong2021prbIV,ZhiDaSong2021prbV,ZhiDaSong2021prbVI}. Strictly speaking, the projection approximation still needs to be improved to consider the renormalization effect from filled bands, like Hartree-Fock approximation (HFA)~\cite{parker2021arxiv,Senthil2023arxiv} or perturbation renormalization~\cite{JianPengLiu2023arxiv,Vafek2020prl}. More precise modeling will be left for future work. In addition, we set $\Delta=1 {\rm meV}$ in Eq.~\eqref{eq:BM} from now on to gap out the lowest two flat bands, which means we will only take conductive band and single valley (K valley) into account for less computation cost. We have numerically checked that when staggered gap is much larger than the single particle flat band width ($\Delta>>W$), the many body energy is not sensitive to staggered gap $\Delta$.

\begin{figure}[htb]
\includegraphics[width=0.9\columnwidth]{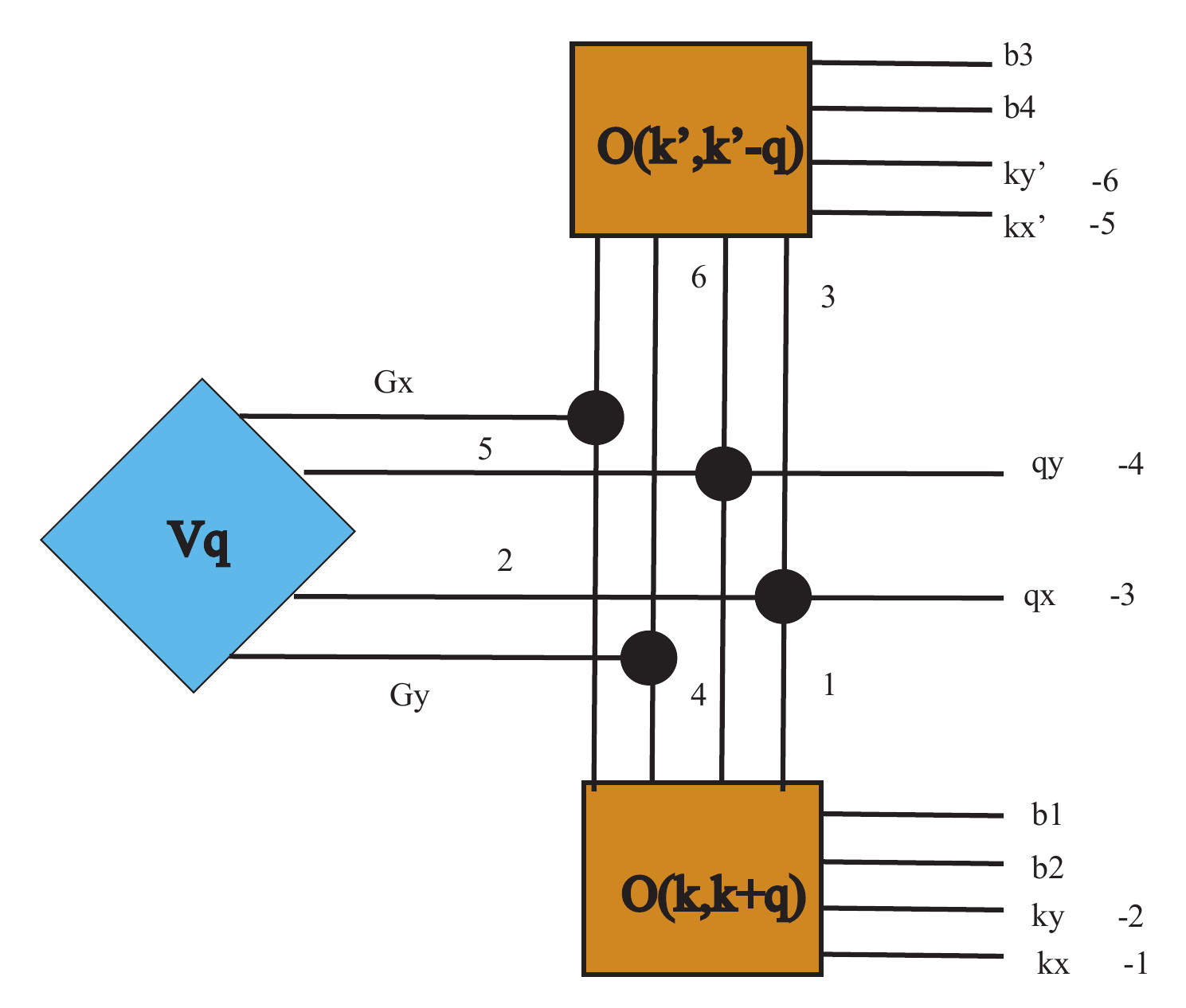}
\caption{The tensor form of the projected interaction matrix element. It can be considered as the contraction between bare Coulomb interaction ($V_{\textbf{q}}$) and flat band form factor($O(\textbf{k},\textbf{k}+\textbf{q})$). Following the convention in tensor network, the legs labelled by negative number are external indices in projected interaction, while legs labelled by positive number are the contracted indices with corresponding ordering. 
The black dots are higher rank kronecker tensor.  For our case, the band indices are all one dimensional so we will omit them in the following. All the contractions are realized by Ncon tensor contractor.}
\label{fig:Vijml}
\end{figure}

The many body Hamiltonian is as follows (Eq.~\eqref{eq:interaction_TBCB}):
 \begin{equation}
	\begin{split}
&H=\sum_{\textbf{k},n}\epsilon_{\textbf{k},n}C^{\dagger}_{\textbf{k},n}C_{\textbf{k},n}+ \frac{1}{2A}\sum_{i,j,k,l,
\textbf{k},\textbf{k}',\textbf{q}}V_{i,j,m,l}(\textbf{k},\textbf{k}',\textbf{q})\\
&(C^{\dagger}_{\textbf{k}+\textbf{q},i}C_{\textbf{k},j}-\bar{\rho}_{\textbf{q}=0}\delta_{i,j})(C^{\dagger}_{\textbf{k}'-\textbf{q},m}C_{\textbf{k}',l}-\bar{\rho}_{\textbf{q}=0}\delta_{m,l}),
 \label{eq:interaction_TBCB}
\end{split}
\end{equation}
where
\begin{equation}
	\begin{split}
&V_{i,j,m,l}(\textbf{k},\textbf{k}',\textbf{q})=\sum_{\textbf{Q}}V(\textbf{q}+\textbf{Q})O_{i,j}(\textbf{k},\textbf{q},\textbf{Q})O_{m,l}(\textbf{k}',-\textbf{q},-\textbf{Q}),
\end{split}
\end{equation}
with
\begin{equation}
	\begin{split}
&O_{i,j}(\textbf{k},\textbf{q},\textbf{Q})= \sum_{\textbf{G},\tau,\mu}u^{*}_{\textbf{k}+\textbf{q}+\textbf{Q},i}(\textbf{G},\tau,\mu)u_{\textbf{k},j}(\textbf{G},\tau,\mu)=\\ &\sum_{\textbf{G},\tau,\mu}u^{*}_{\textbf{k}+\textbf{q},i}(\textbf{G}-\textbf{Q},\tau,\mu)u_{\textbf{k},j}(\textbf{G},\tau,\mu).
\end{split}
\end{equation}

\begin{figure}[htb]
\includegraphics[width=0.98\columnwidth]{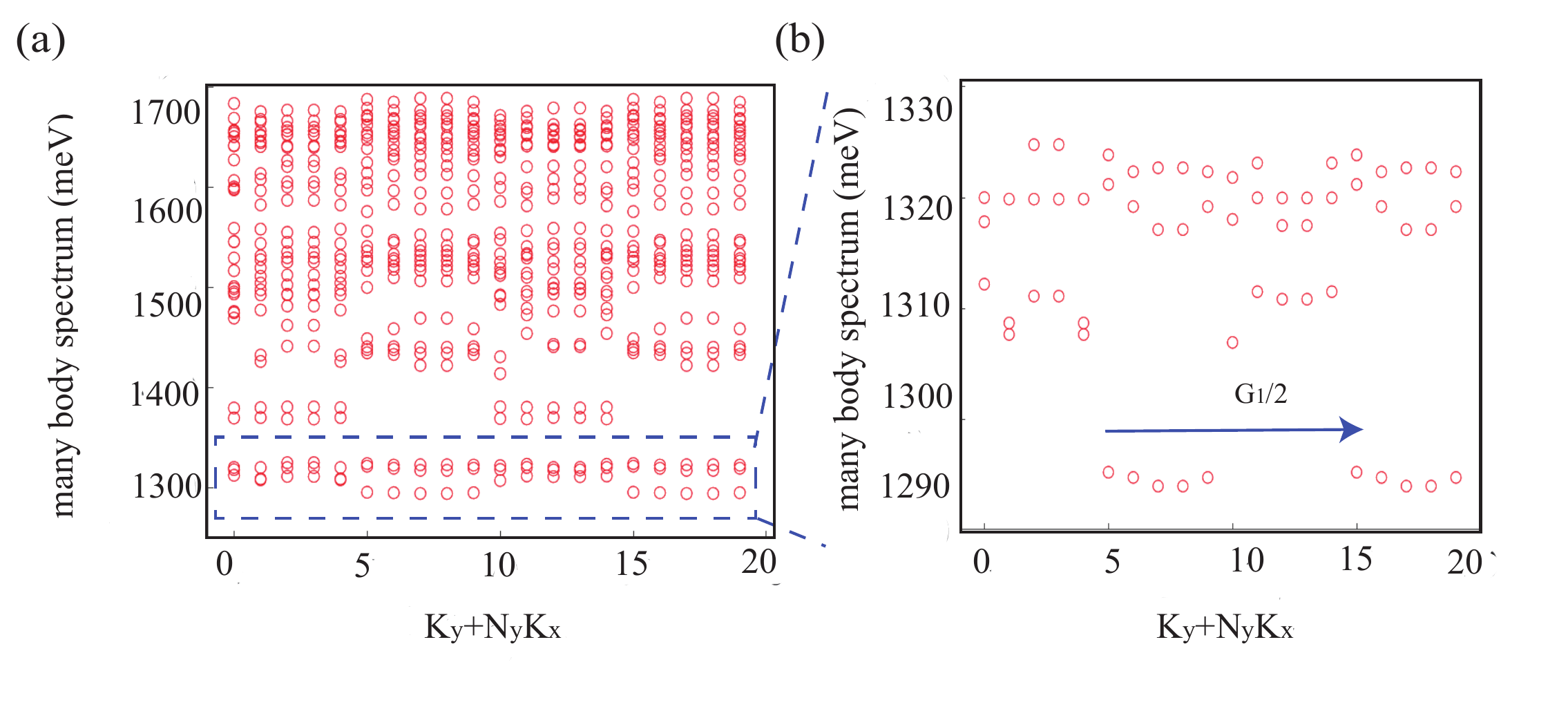}
\caption{(a) The many body energy spectrum for $\nu=1/5$ MATBCB under chiral limit. The discretization of the first mBZ is $4\times 5$. The number of mBZ is $7\times 7$ which is the size of quadratic nodes array. We will use the same parameters in the following simulation unless specified explaination. 
 (b) The zoomed in of low energy window in (a). One can identify the ground state manifold with GSD=10 at total momentum sector $K=5,6\cdots 9$ and $K=15,16\cdots 19$. They are approximately related by vector $G_{1}/2$. The gap between ground state manifold and the higher excitation states is about 12 meV.}
	\label{fig:ED_energy}
\end{figure}

Here $\epsilon_{\textbf{k},n}$ represents the conductive band single particle energy given by Hamiltonian Eq.~\eqref{eq:BM}. This kinetic term will be overwhelmed by interaction near magic angle. The flat band will have well defined Chern number $C=2$ after gapping out. $C_{\textbf{k},n}$ is the flat band fermion annihilation operator, $\textbf{k}$ labels momentum and n labels band. $A=N_{x}N_{y}$ is the normalized area which corresponds to the k point discretization of the first mBZ.
In the following, we use the discretization choice of momentum:
\[
k_{x}=\frac{2\pi}{N_{x}}(0,1\cdots ,N_{x}-1), k_{y}=\frac{2\pi}{N_{y}}(0,1\cdots ,N_{y}-1).
\]
$V_{i,j,m,l}(\textbf{k},\textbf{k}',\textbf{q})$ represents the matrix elements of the projected Coulomb interaction. Since we only consider conductive band, the band indices $i,j,m,l$ are one dimensional and we will omit them after that. The tensor form of $V_{i,j,m,l}(\textbf{k},\textbf{k}',\textbf{q})$ is shown in Fig.~\ref{fig:Vijml}. The concrete matrix elements are computed by ncon tensor contractor~\cite{Vidal2015arxiv}. $V(\textbf{q})=\frac{2\pi e^{2}}{\epsilon_{r}\epsilon_{0}}\frac{\tanh(|\textbf{q}|d/2)}{|\textbf{q}|}$ is the dual gate bare screened Coulomb potential in momentum space. d is the dual gate distance and $\epsilon_{r}$ is the relative dielectric constant. In principle it can be generalized to layer resolved version~\cite{JianKang2020prb}, here for simplification we do not resolve the layer in $V(\textbf{q})$. $O_{i,j}(\textbf{k},\textbf{q},\textbf{Q})$ is the flat band projective form factor. The overlap sums over quadratic reciprocal points $\textbf{G}$, sublattice $\tau$, layer $\mu$. 
The analytical form of the form factor has rich algebraic structure~\cite{ZhiDaSong2021prbIV,HongYao2022prr}. Numerically, there is phase ambiguity for Bloch function at each k point. So one has to fix the gauge (the phase difference between each pair of k points) to make projective interaction matrix element smooth about $\textbf{k},\textbf{k}',\textbf{q}$. Due to the intrinsic topological non-trivial flat band, the phase discontinuity is unavoidable~\cite{Zaletel2020prb,Regnault2014prb,QiMiaoSi2024arxiv}. One can hide it at the boudnary of the first mBZ. The phase of form factor is fixed to $O(\textbf{k},\textbf{q},\textbf{Q})\to |O|\exp[i\frac{2\pi C}{G_{1}G_{2}}(q_{x}+Q_{x})(q_{y}+Q_{y})]$. When $\textbf{k}+\textbf{q}$ goes beyond the first mBZ, one has to pull back it to the first mBZ with reciprocal vector $\tilde{G}$ and compensate it in the quadratic node array as Fig.~\ref{fig:singlebody_band}.

\begin{figure}[htb]
\includegraphics[width=0.98\columnwidth]{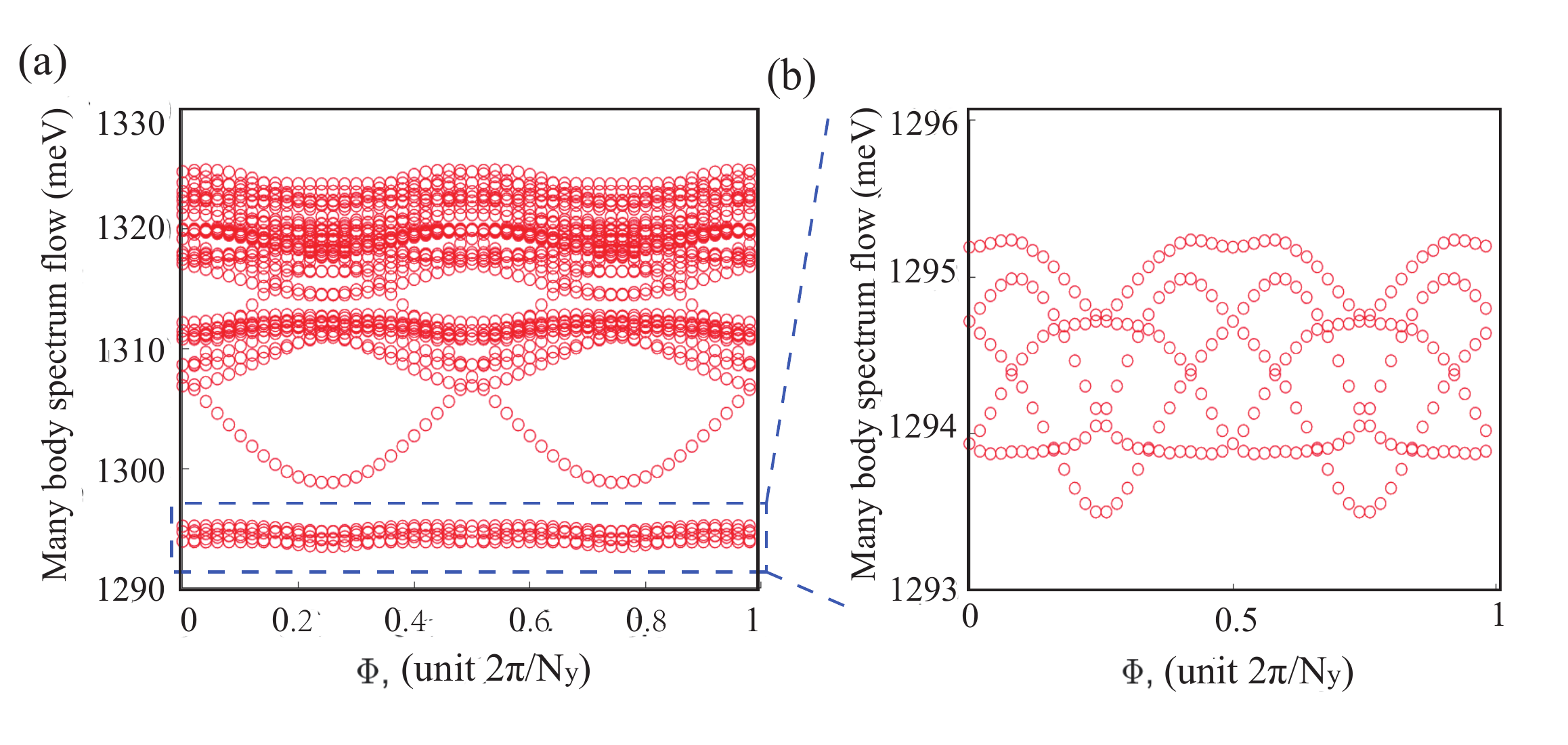}
\caption{(a) The spectrum flow under adiabatic flux insertion $k_{y}\to k_{y}+\Phi/N_{y}$ under chiral limit. (b) The zoomed in of low energy window in (a). One can examine that the states related by $G_{1}/2$ will keep almost two-fold degeneracy during flux inserting. The ten-fold quasi-degenerate ground states braid with each other as flux varying, while do not mix with higher energy states. These 10 states can be considered as the FCI topological ground state candidates. The gap between ground state manifold and higher energy states is about $5-12 {\rm meV}$, While the width of ground state manifold is approximately $1-2 {\rm meV}$. 
}
	\label{fig:spectrum_flow}
\end{figure}

Unlike the integer filling situation whose ground state is almost a Slater determinant~\cite{ZhiDaSong2021prbIV,HongYao2022prr}, for such fractional filling system, the ground state generally is the linear combination of Slater determinants. There is usually a serious sign problem for fractional filling. That means one can not naively use determinant quantum Monte-Carlo (DQMC) algorithm~\cite{GaoPeiPan2024ECMP,XuZhang2022prb} to model MATBCB. To realize the sign problem free simulation of Hamiltonian Eq.~\eqref{eq:interaction_TBCB}, like using exact diagonalization (ED), one needs to make use of symmetry to reduce the matrix block dimension, e.g. the particle number conservation. The particle number is defined by the filling factor $\nu=\frac{N_{e}}{N_{x}N_{y}}$. According to the vortexability formalism~\cite{Patrick2022arxiv,junkaidong2023prr}, for $\nu=1/5$, $C=2$ system with contact interaction, it is expected to host a translation symmetry reserved FCI phase. So we will focus on $\nu=1/5$ in this manuscript. In addition, one can use tranlation symmetry to consider smaller block, where each block or K sector is labeled by total momentum~\cite{Sandvik2010lecture}. Total momentum will be restricted in the first mBZ, i.e., $K_{i}\to {\rm mod}(K_{i},N_{i}), i=x,y$.

Empirically, we expect relatively strong interaction would stablize FCI phase. So we choose a relative large gate distance $d=1000a\approx 36 a_{M}$ and small dielectric constant $\epsilon_{r}=1$ in following simulation. Here $a$ is the monolayer lattice constant while $a_{M}$ is the moir\'e superlattice constant. 

Fig.~\ref{fig:ED_energy} shows the many-body energy spectrum for $\nu=1/5$ MATBCB system in the chiral limit with size $4\times 5$ k points. We will use the same parameters in the following simulation unless otherwise specified. It can be seen that there is a ten-fold quasi-degenerate ground state. The gap between ground state manifold and higher excitation states is about $10-12 {\rm meV}$. The ground state at $K=5,6\cdots 9$ and $K=15,16\cdots 19$ sector can be approximately related by $G_{1}/2$ vector. Later we will show that additionally, the lowest ground state at different K sector can also be related by $G_{2}/2$ from the spectrum flow under adiabatic flux inserting. One can conjecture that in putative topological degenerate ground state manifold, the ground state degeneracy(GSD) is 10. However, one should keep in mind that up to now we can not claim there is FCI phase in $\nu=1/5$ MATBCB. We still have to inspect the ground state spectrum flow under adiabatic flux inserting~\cite{YiFengWang2023prb,Bernevig2011prx}. The ground states should keep in the low energy subspace without gap closing with other higher energy states, which means the GSD should be stable under the adiabatic variation of boundary condition. 

In Fig.~\ref{fig:spectrum_flow}, we show the corresponding spectrum flow for the same size and filling. We adapt the periodic boundary condition in $k_{y}$ direction and let the momentum occur global shift $k_{y}\to k_{y}+\Phi/N_{y}$. It can be considered as one rolls $k_{y}$ direction as a cylinder and inserts flux adiabatically along the cylinder axis. One can see the putative GSD=10 is stable in spectrum flow. The almost two-fold degeneracy for low energy state related by $G_{1}/2$ is preserved during flux inserting.  When $\Phi\in [0,2\pi/N_{y})$, one can see that there is a non-trivial braiding between quasi-degenerate ground states. Moreover, in Fig.~\ref{fig:spectrum_flow} one can observe that spectrum flow is symmetric about $\Phi_{0}=\pi/N_{y}=\pi/5$. That means the lowest many-body spectrum is also invariant under the translation $G_{2}/2$. This indicates in many-body level, there may be similar folding rule along $\textbf{Q}=(G_{1}/2,G_{2}/2)$ like single particle counterpart~\cite{Patrick2022prl,junkaidong2023prr}. In simulation, one should be cautious that keeping aspect ratio to be compatible is important. Otherwise the simulation may not approach 2d thermodynamic limit and spurious energy gap may emerge ~\cite{Bernevig2011prx}. One can notice that when $\Phi$ changes from 0 to $2\pi/N_{y}$, $N_{y}$ of $k_{y}$ cuts exchange with each other. However their relative relationship in momentum space is unchanged. So $2\pi/N_{y}$ is one of the spectrum flow period. 

In addition, one can also consider the many body Chern number~\cite{Barkeshli2021prb,WeiZhu2024arxiv,BinBinChen2024arxiv} of the ten-fold degenerate states under chiral limit. By adiabatically inserting the flux along both direction of torus, one can get $C_{many}=\frac{1}{2\pi}\int_{0}^{2\pi}d\theta_{x}\int_{0}^{2\pi}d\theta_{y}F(\theta_{x},\theta_{y})$. It is also convenient to use the similar formalism like Eq.~\ref{eq:Wilson_LGT} in an effective discrete flux lattice. We get the total many body Chern number $|C_{tot}|=4$. The many body Chern number $|C_{many}|=2/5$ is the average of $C_{tot}$ over ten-fold degenerate ground state, which is directly related to many body Hall conductivity. The many-body Chern number $|C_{many}|$ can be related to the Chern number of single particle band $|C_{single}|$ via $|C_{many}|=|C_{single}|\nu$ ($\nu$ is the filling fraction)~\cite{DNshengLFu2024arxiv}, in consistence with former work in the Hofstadter model~\cite{Neupert2021prb}. Since the $C_{many}$ is a fractional one, it has ruled out some symmetry breaking Chern insulator phase like quantum anomalous Hall crystal~\cite{DNshengLFu2024arxiv}.

Finally in this section, even if the spectrum flow shows that there could be topologically degenerate states in $\nu=1/5$ MATBCB, one still can not distinguish FCI, potential commensurate charge density wave (CDW) and other competing phases~\cite{Emil2024arxiv}. To get the smoking gun evidence of FCI phase, one still has to diagnose the topological nature of many body ground state like many body entanglement. In the next section, we will settle down the FCI phase via particle entanglement spectrum (PES) and study the phase transition induced by chiral ratio variation through monitoring the PES gap.   

\section{The particle entanglement spectrum for MATBCB and quasi-hole counting}
\label{sec:PES}

\begin{figure}[h]
\includegraphics[width=1\columnwidth]{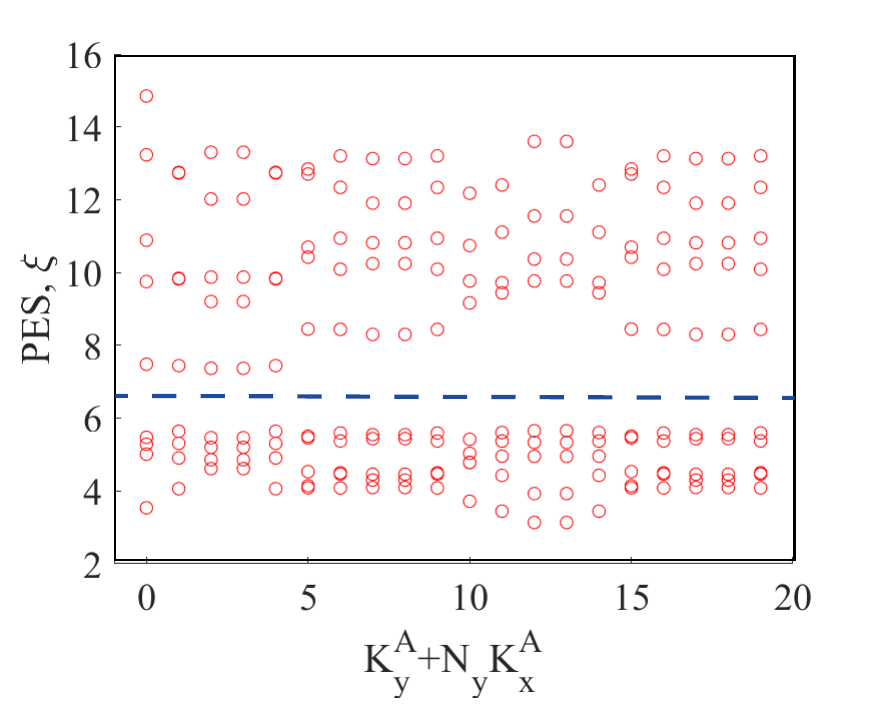}
\caption{The PES of ten-fold ground state density matrix for chiral limit $\kappa=w_{AA}/w_{AB}=0$. The PES is labelled by the total momentum of subsystem A $K^{A}$. The spectrum almost gather at $\xi<16$ regime, while other data with higher values can be considered as noise. 
There is an apparent PES gap at $\xi\approx 6.5$. One can extract the PES quasi-hole counting at each $K^{A}$ (total momentum of subsystem A) sector under the PES gap. For $K^{A}=0,1\cdots 4$, there are 4 states per sector. While for $K^{A}=5,6\cdots 19$, there are 5 states per sector. The total number of quasi-hole is 95 (for $4\times 5$ momentum space size). The counting obeys the $(1,5)_{C=2}$ GPP based on Halperin spin singlet ansatz as argued in main text.}
\label{fig:PES}
\end{figure}

To reveal the exotic quasi-hole excitation in putative FCI phase, one can consider the particle entanglement spectrum (PES)~\cite{Haldane2008prl} for ground state or ground state manifold. The quasi-hole counting for FCI phase will obey generalized Pauli principle (GPP) and depend on the topology of base manifold~\cite{Bernevig2011prx,Regnault2013prb}. So the PES quasi-hole counting will be a smoking gun evidence of FCI phase and help one rule out CDW phase or other topologically trivial competing phases. Recently, PES has been widely used in FCI or FQAHE candidates from Moir\'e system~\cite{LiangFu2024prbletter,ZhaoLiu2024arxiv,KaiSun2024arxiv,ZhaoLiu2022prl} and R5G system~\cite{Senthil2023arxiv}. Briefly speaking, PES is the eigenvalue of the reduced density matrix which bipartites the system in particle number (Fock) space. The mostly used reduced density matrix is to trace over the all possible half of the particle configurations in subsystem B in the model and keep the other half subsystem A~\cite{Haldane2008prl}. However, the particle number of subsystem A and B are not necessarily to be the same.

Following Ref.~\cite{Schoutens2009JPMT}, one can consider the pure state PES for simplicity. The ground state of fractional filling MATBCB generally is a linear combination of Slater determinants.  If the ground state is non-degenerate, one can write down the ground state as follow, (which is a pure state):
\begin{equation}
\begin{split}
 &|\psi_{0}\rangle =\left( \begin{array}{ccc} a_{1}  \\ a_{2} \\ \cdots \\
 a_{n}
 \end{array}\right)=a_{1}\left( \begin{array}{ccc} 1  \\ 0 \\ \cdots \\
 0
 \end{array}\right)+\cdots +a_{n}\left( \begin{array}{ccc} 0  \\ 0 \\ \cdots \\
 1
 \end{array}\right)=a_{1}\phi_{1}+\cdots+a_{n}\phi_{n}\\
 &=\frac{1}{\sqrt{C(N_{e},N_{A})}}[a_{1}(\sum_{A,B}(-1)^{sign_{1}(A,B)}\phi_{1}^{A}\phi_{1}^{B})+\cdots+\\
 &a_{n}(\sum_{A,B}(-1)^{sign_{n}(A,B)}\phi_{n}^{A}\phi_{n}^{B})].
 \end{split}
 \end{equation}

\begin{figure*}[htb]
\includegraphics[width=2\columnwidth]{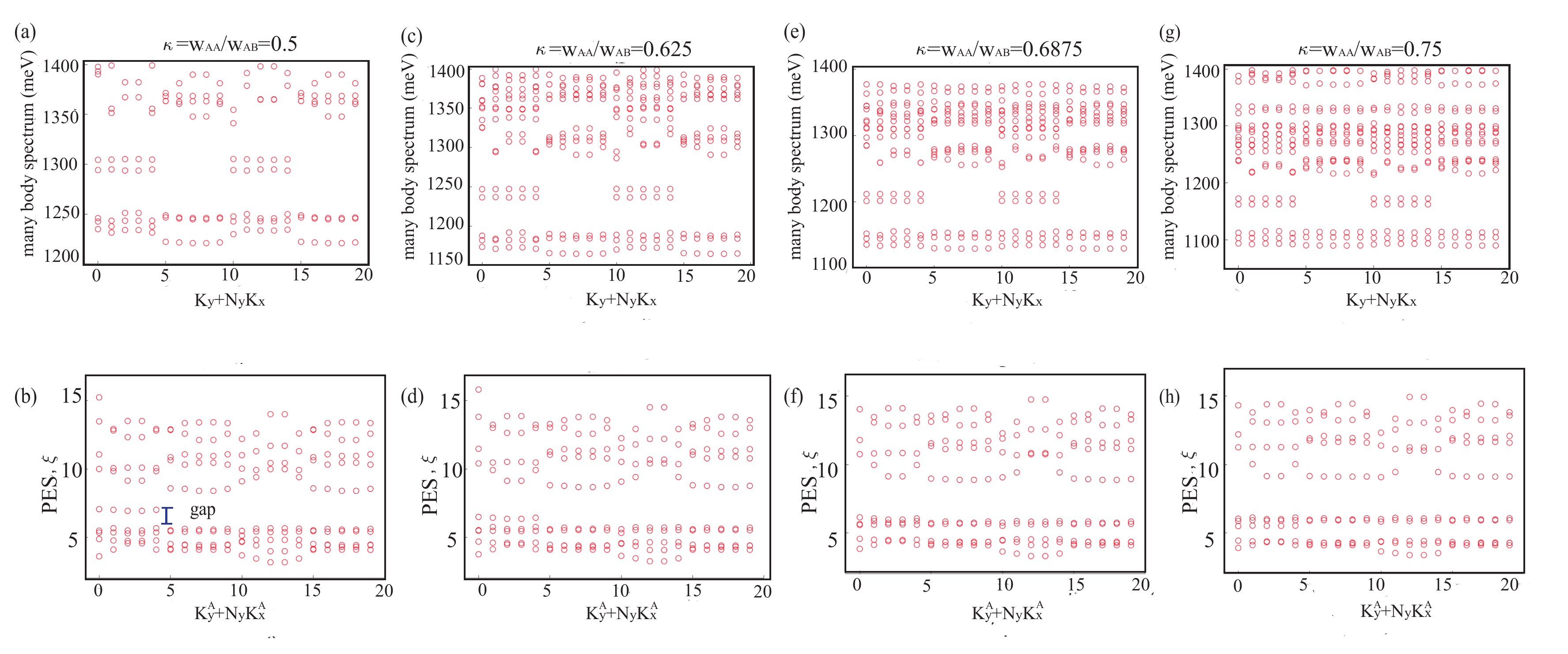}
\caption{(a) The many-body energy spectrum of MATBCB for $\kappa=w_{AA}/w_{AB}=0.5$. One can see that the original ten-fold ground states get closer to higher energy states. (b) The PES of 10 fold degenerate ground state density matrix for the same $\kappa$. The PES gap is illustrated by the blue line segment. The location of PES gap is the same in (d), (f), and (h). The quasi-hole counting under PES gap is the same as $\kappa=0$ case, which indicates FCI phase is stable under the variation of chiral ratio to some extend. There are five states above the PES gap which locate at $\xi\approx 7$ and $K^{A}=0,1\cdots 4$. Compared with $\kappa=0$ case, they are approaching the original quasi-hole manifold when $\kappa$ increases to $0.5$. So we expect when $\kappa$ increases to a critical value, these five higher states will meet the lower quasi-hole states. (c) The many-body energy spectrum of MATBCB for $\kappa=0.625$. (d) The PES for $\kappa=0.625$. The five states mentioned above get very close to lower quasi-hole states. (e) The many-body energy spectrum of MATBCB for $\kappa=0.6875$. (f) The PES for $\kappa=0.6875$. The five states mentioned above have partially collapsed to lower quasi-hole states. (g) The many-body energy spectrum of MATBCB for $\kappa=0.75$. (h) The PES for $\kappa=0.75$. The five states mentioned above have completely collapsed to lower quasi-hole states. This may be interpreted as the violation of the hard core condition. After phase transition, the particle with different pseudo spin can occupy the same orbital. So the number of lower quasi-hole states increase extra $\frac{20}{2^{2}}=5$.}
\label{fig:ED_PES_kappa}
\end{figure*}

Here ${\phi_{1},\cdots,\phi_{n}}$ are Slater determinants in the first quantization form. By using the famous Laplace decomposition formalism for determinant in linear algebra, one can expand each Slater determinant with A and B sub-determinants, the sign like $sign_{1}(A,B)$ is determined by the sign of the algebraic cofactor. $\frac{1}{\sqrt{C(N_{e},N_{A})}}=\frac{1}{\sqrt{(N_{e}!)/[(N_{A})!(N_{B})!]}}$ occurred in prefactor is for normalization. In principle one can write out the bipartition of wave function $|\psi_{0}\rangle$ under the particle conserved basis with particle number $N_{A}=N_{e}/2$. Its dimension is $C(N,N_{A})=N!/[N_{A}!(N-N_{A})!]$. Here $N=N_{x}N_{y}$ is the number of total orbital. The nonzero singular values of matrix $|\psi_{0}\rangle_{C(N,N_{A})\times C(N,N_{B})}$ are denoted as $\{\exp(-\xi_{i}/2)\}$, where $\{\xi_{i}\}$ is the PES.

However, the above procedure only applies to the pure state or non-degenerate state. The situation for mixed state or degenerate state in our $\nu=1/5$ MATBCB system will be more complicated. One has to consider the following mixed state density matrix consisting of degeneratestates~\cite{Bernevig2011prx,Senthil2023arxiv}.
  \begin{equation}
	\begin{split}
 \rho=\frac{1}{D}\sum_{i=1}^{D}|\psi_{i}\rangle \langle\psi_{i}|.
  \end{split}\label{eq:density matrix}
 \end{equation}

Generally speaking, each term in Eq.~\eqref{eq:density matrix} will not share common A, B bipartitions. However, one can still find a large enough subspace (such as the $C(N,N_{A})\times C(N,N_{B})$ dimension space mentioned above.) to trace over the all possible B part term by term in Eq.~\ref{eq:density matrix}, which means one can always do the SVD term by term and then get the reduced density matrix $\rho_{A}$. The eigenvalues of $\rho_{A}$ will be 
 $\{\exp(-\xi_{i})\}$ which is related to PES $\{\xi_{i}\}$. 
\begin{equation}
\begin{split}
 &|\psi_{i}\rangle=U_{i}S_{i}V_{i}^{\dagger}, \\
 &\rho_{A}=Tr_{B}[\rho]=\frac{1}{D}\sum_{i=1}^{D}Tr_{B}[|\psi_{i}\rangle \langle\psi_{i}|] 
 = \frac{1}{D}\sum_{i=1}^{D}U_{i}S^{2}_{i}U_{i}^{\dagger}.
\end{split}
\label{eq:reduced density matrix}
\end{equation}

After introducing the algorithm of PES, in order to physically interpret the result of PES, we now briefly review the generalized Pauli principle (GPP) and its higher Chern number generalization~\cite{Bernevig2011prx,Regnault2013prb}. For the simplest $C=1$ case, one has $(k,r)_{C=1}$ GPP in $\nu=k/r$ FCI phase for its quasi-hole counting. Concretely, in Fock space, there should be no more than k occupied states among arbitrary consecutive r orbitals. For FCI phase, the number of quasi-hole state under PES gap should be consistent with the configuration number predicted by GPP. The qusi-hole counting in GPP will be sensitive to the base manifold topology. For example, on sphere the orbitals align in a chain. However for our case on $T^{2}$ (torus), the orbitals align in a ring which means the occupation at the end should also obey the $(k,r)_{C=1}$ constraint. For $\nu=1/3$, there is an analytical solution for $(1,3)_{C=1}$ GPP configuration number $N_{x}N_{y}\frac{(N_{x}N_{y}-2N_{A}-1)!}{(N_{x}N_{y}-3N_{A})!(N_{A}!)}$~\cite{Bernevig2011prx}, which corresponds to Laughlin state quasi-hole counting. Unfortunately, for the case of general $(k,r)_{C=1}$, the GPP quasi-hole counting is related to so called Jack polynomial~\cite{bernevig2008prl,AiLeiHe2019prb}, for which there is no simple form in general. On the other hand, numerically one can always enumerate all configurations obeying $(k,r)_{C=1}$ GPP on torus for not so large size system. For higher Chern number $C>1$ situation, it is believed that the quasi-particle will carry an additional internal SU(C) degree of freedom, which bring more constraints. This SU(C) symmetry is usually called color symmetry like high energy community. Specially, when $C=2$, the SU(2) symmetry can be considered as an effective (pseudo) spin degree of freedom. During bipartition, it is reasonable to assume that the internal SU(C) symmetry is preserved in each subsystem A and B. That means in subsystem A the quasi-particle should form (Halperin) color singlet or minimize the net color polarization~\cite{Regnault2013prb}. For the particle with the same color, they will obey the $(k,r)_{C=1}$ GPP. While for the counterpart with different color they will not have such constraint except the hard core condition. That means two particles can not occupy the same orbital even if they have different color. An interpretable analogy for $C=2$ case is the spinful Pauli exclusive principle under strong intra orbital interaction limit in atomic physics. The particle with different spin will not occupy the same orbital.

\begin{figure}[t]
\includegraphics[width=0.98\columnwidth]{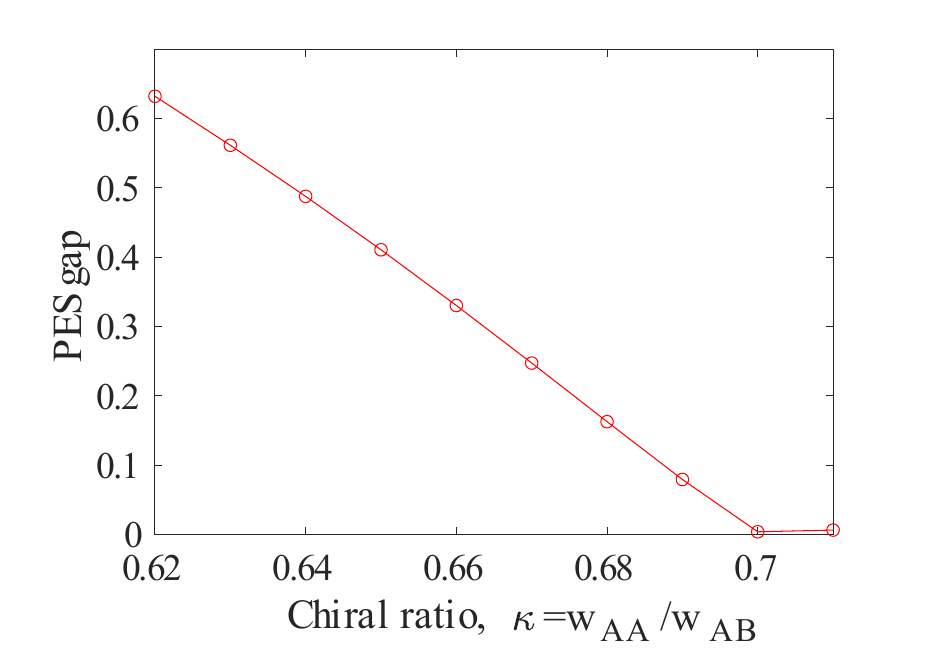}
\caption{The PES gap versus chiral ratio. The PES gap is defined as $|\min(\xi(n=5,K^{A}=0\cdots 4))-\max(\xi(n=4,K^{A}=0\cdots 4), \xi(n=5,K^{A}=5\cdots 19))|$. The gap decreases with increasing chiral ratio and becomes close to zero at $\kappa_{C}\approx 0.7050$, suggesting a phase transtion.} 
\label{fig:PESgap}
\end{figure}

From the above analysis, now one can come back and inspect the PES of $\nu=1/5$ MATBCB. In Fig.~\ref{fig:PES}, we show the PES of the density matrix consists of 10 fold topological ground state candidates. There is an apparent PES gap at about $\xi=6.5$ as shown by the blue dashed line. One can check that there are 4 quasi-hole states under PES gap in each $K^{A}=0,1\cdots 4$ sector while 5 low quasi-hole states in each $K^{A}=5,6\cdots 19$ sector. The total number of quasi-hole state is 95 for $N_{x}=4, N_{y}=5, N_{A}=2$. Such quasi-hole counting is coincident with $(1,5)_{C=2}$ GPP. One can interpret this result as follows.  In our case, $N_{x}=4, N_{y}=5, N_{e}=4, N_{A}=2$. One can enumerate the Halperin spin singlet in subsystem A. The possibility to get spin up or down particle are the same $1/2$ (1/C). So the number of configurations which obeys $(1,5)_{C=2}$ GPP is $\frac{20\times 19}{2^{2}}=95$. While for CDW phase, the quasi-hole counting is significantly different, which is given by $n=d C(N_{e},N_{A})$~\cite{BernevigRegnault2012arxiv,Emil2024arxiv}. $d$ is the number of CDW order. Up to now, one can claim that $\nu=1/5$ MATBCB under chiral limit hosts $C=2$ abelian FCI phase. MATBCB can naturally realize the $C=2$ FCI without stacking $C=1$ FCI. For the later one, it is hard to guarantee the position of the valley is invariant when stacking. 

On the other hand, it is important to check if such $C=2$ FCI is robust when it deviates from the ideal condition~\cite{Oreg2024arxiv}. For example, the influence of chiral ratio $\kappa=w_{AA}/w_{AB}$~\cite{Zaletel2020prb,JianKang2020prb,parker2021arxiv}. When $\kappa\neq 0$, the lowest single particle bands will be broadened. The kinetic term will compete with projective interaction. And quasi-particle will generally feel a non-uniform Berry curvature. One can expect there is a critical chiral ratio where the FCI phase breaks down and phase transition happens. We will locate such phase transition according to PES gap~\cite{Bernevig2011prx}. In Fig.~\ref{fig:ED_PES_kappa}, we show the many-body energy spectrum and the PES of 10 fold degenerate mixed state under different chiral ratio $\kappa=0.5, 0.625, 0.6875, 0.75$ while keeping other parameters unchanged. When $\kappa$ increases, there are 5 states approaching the original quasi-hole manifold at $\kappa=0$. As $\kappa$ reaches the critical value $\kappa_{c}$, these 5 states will collapse on lower quasi-hole manifold.
Denoting $|\min(\xi(n=5,K^{A}=0\cdots 4))-\max(\xi(n=4,K^{A}=0\cdots 4), \xi(n=5,K^{A}=5\cdots 19))|$ as the PES gap, the PES gap closing phase transition can be interpreted as the violation of hard core condition. That means after phase transition, the particle with different pseudo spin can occupy the same orbital. So one has additional $\frac{20}{2^{2}}=5$ configurations. It is not difficult to understand this since when $\kappa$ increases, the kinetic term can be comparable with interaction at a critical value. By keeping track of the PES gap evolution about chiral ratio $\kappa$, we find that the PES gap decreases as $\kappa$ increases, shown in Fig.~\ref{fig:PESgap}. And there is a PES gap closing at $\kappa_{C}\approx 0.7050$, which indicates a phase transition.

To summarize this section, we have obtained the smoking gun evidence of the existence of $C=2$ FCI phase in $\nu=1/5, \kappa=0$ MATBCB via PES quasi-hole counting. The number of quasi-hole states under PES gap obeys $(1,5)_{C=2}$ GPP. As chiral ratio $\kappa$ increases, there will be PES gap closing at $\kappa_{c}\approx 0.7050$. The quasi-hole counting will change, which indicates the violation of the hard core condition. However, recall that in section ~\ref{sec:ED_and_spec_flow} we get the spectrum flow with about $5-12 {\rm meV}$ gap between ground state manifold and higher energy states. This gap is not large enough compared with the ground state manifold width $\approx 1.5 {\rm meV}$. One may expect a weak external magnetic field or exchange interaction~\cite{YongXu2024arxiv} is still needed to stabilize the FCI phase~\cite{YongLongXie2021Nature,Zaletel2024arxiv,Andrews2018prb,Neupert2021prb}, which will be left for future work.
		   
		   

\section{Conclusion and discussion}
\label{sec:disc}

In this manuscript, we study a putative $C=2$ abelian FCI phase in $\nu=1/5, \kappa=0$ MATBCB. In the single particle level, we obtain the lowest topological flat bands with $C=\pm 2$. The band width is about $1.5\times 10^{-3} {\rm meV}$ at the first magic angle $\phi=1.608^{\circ}$. The almost uniform Berry curvature and tiny trace condition violation indicate the ideal quantum geometry condition to host FCI phase. In the many-body level, we first present the many body energy spectrum and spectrum flow for projected Coulomb interaction model of MATBCB. We find there is topological ground state manifold candidate with GSD=10. Next we examine the PES quasi-hole counting of the ground state manifold candidate. We find the counting meets the $(1,5)_{C=2}$ GPP as Halperin spin singlet. This counting is a smoking gun evidence of $C=2$ FCI. When the parameters are not ideal, such as the chiral ratio deviates from zero, there will be a phase transition which breaks the FCI phase. We estimate that $\kappa_{c}\approx 0.7050$. More comprehensive research on the FCI relevant phase transition will be left for future work.

However, one should still keep in mind that the results and arguments above are all built on the assumption that the translation symmetry is preserved. When the translation symmetry is broken, the internal SU(C) color symmetry can be entangled with spatial symmetry. Especially, the SU(C) symmetry and spatial translation symmetry will combine to form a generalized (pseudo) magnetic translation symmetry~\cite{Regnault2014prb,YingHaiWu2015prb}. In that case, we conjecture that the SU(C) color symmetry will also be broken accompanied with the translation symmetry broken. An example is the $\nu=1/6, C=2$ flat band system~\cite{junkaidong2023prr}. At that situation, Halperin color singlet quasi-hole counting ansatz based $(k,r)_{C}$ GPP may be violated. A more general pseudo potential method under $n_{x}-k_{y}$ basis has to be used~\cite{Regnault2014prb,XiaoLiangQi2011prl,XiaoLiangQi2013prb}. Moreover, under $n_{x}-k_{y}$ basis it is convenient to simulate fractional filling system with numerical methods like DMRG~\cite{Zaletel2020prb,JianKang2020prb}, which can handle larger size compared with ED. Compared with analytical pseudopotential method, it can model the situation beyond thin cylinder limit~\cite{Regnault2014prb}. It is also very interesting and valuable to search for possible non-abelian FCI phase in TBCB system. The problem about the coexistence of FCI phase and charge ordered phase~\cite{BinBinChen2024arxiv} is also beyond one's intuitive knowledge, which may occur in TBCB. Also, it is worth to find the connection between higher Chern number FCI and topological quantum chemistry in order to find material realization~\cite{Yunzheliu2024arxiv}. For spinful case there may be much richer competing phases, e.g., ferromagnetic phase. Also it is very interesting to study the spin-valley locking effect and the compitition between FCI and valley-polarized phase like $\rm{MoTe_{2}}$ ~\cite{DiXiao2024prl} or spin-valley composite phase in moir\'e trilayer graphene~\cite{GuoYiZhu2018arxiv,GuoYiZhu2018SciBulletin} case. The open questions mentioned above are worth further research in the future.

\section*{Acknowledgements}
We thank Jian Kang, Wei Li, Daniel E.Parker, Han-Qing Wu, Jie Wang, Ying-Hai Wu, Jia-Qi Cai, Gao-Pei Pan, Ming-Rui Li, Jun-Kai Dong, Ti-Xuan Tan, Zhongqing Guo, Wang-Qian Miao, Yunzhe Liu, Xiaohan Wan, Mu-Wei Wu for valuable discussions. J. Z. M. and D. X. Y. are supported by National Key Research and Development Program of China (2022YFA1402802), National Natural Science Foundation of China (92165204), Leading Talent Program of Guangdong Special Projects (201626003), Guangdong Provincial Key Laboratory of Magnetoelectric Physics and Devices (Grant No. 656
2022B1212010008), Guangdong Fundamental Research Center for Magnetoelectric Physics, and Shenzhen International Quantum Academy. J.-Y. C. is supported by  National Natural Science Foundation of China (12304186), Open Research Fund Program of the State Key Laboratory of Low-Dimensional Quantum Physics (KF202207), Fundamental Research Funds for the Central Universities, Sun Yat-sen University (23qnpy60), Innovation Program for Quantum Science and Technology 2021ZD0302100, Guangzhou Basic and Applied Basic Research Foundation (2024A04J4264), the
Guangdong Basic and Applied Basic Research Foundation (2024A1515013065). R.Z.H is supported by a postdoctoral fellowship from the Special Research Fund (BOF) of Ghent University.

\bibliography{TBCB}
\end{document}